\documentclass[12pt,a4paper]{article}
\pdfoutput=1
{\def\usepackage{ws-procs9x6}}

\ifnum\pdfoutput=1\else
\PassOptionsToPackage{hypertex}{hyperref}
\PassOptionsToPackage{draft}{graphicx}
\usepackage{showkeys}
\fi

\usepackage{amsmath,amssymb}
\usepackage[bookmarks=true,hyperfigures=true]{hyperref}
\usepackage{graphicx}
\usepackage[nosort]{cite}
\usepackage[bulletsep]{collref}

\usepackage[a4paper,text={450pt,650pt},centering]{geometry}

\let\oldbfseries=\bfseries
\let\oldmdseries=\mdseries
\let\oldnormalfont=\normalfont
\renewcommand{\bfseries}{\oldbfseries\boldmath}
\renewcommand{\mdseries}{\oldmdseries\unboldmath}
\renewcommand{\normalfont}{\oldnormalfont\unboldmath}

\allowdisplaybreaks[3]

\numberwithin{equation}{section}

\usepackage[font=small,labelfont=bf,width=0.85\textwidth]{caption}

\providecommand{\hypersetup}[1]{}
\providecommand{\texorpdfstring}[2]{#1}

\hypersetup{plainpages=false}
\hypersetup{pdfpagemode=UseNone}
\hypersetup{bookmarksnumbered=true}
\hypersetup{pdfstartview=FitH}
\hypersetup{colorlinks=false}
\hypersetup{citebordercolor={.5 1 .5}}
\hypersetup{urlbordercolor={.5 1 1}}
\hypersetup{linkbordercolor={1 .7 .7}}

\DeclareMathSymbol{\Gamma}{\mathalpha}{letters}{"00}
\DeclareMathSymbol{\Delta}{\mathalpha}{letters}{"01}
\DeclareMathSymbol{\Theta}{\mathalpha}{letters}{"02}
\DeclareMathSymbol{\Lambda}{\mathalpha}{letters}{"03}
\DeclareMathSymbol{\Xi}{\mathalpha}{letters}{"04}
\DeclareMathSymbol{\Pi}{\mathalpha}{letters}{"05}
\DeclareMathSymbol{\Sigma}{\mathalpha}{letters}{"06}
\DeclareMathSymbol{\Upsilon}{\mathalpha}{letters}{"07}
\DeclareMathSymbol{\Phi}{\mathalpha}{letters}{"08}
\DeclareMathSymbol{\Psi}{\mathalpha}{letters}{"09}
\DeclareMathSymbol{\Omega}{\mathalpha}{letters}{"0A}

\newcommand{\gen}[1]{\mathrm{#1}}
\newcommand{\superN}{\mathcal{N}}

\newcommand{\Tr}{\mathop{\mathrm{Tr}}}

\newcommand{\Complex}{\mathbb{C}}
\newcommand{\Reals}{\mathbb{R}}
\newcommand{\Projective}{\mathbb{P}}
\newcommand{\Mink}{\mathbb{M}}
\newcommand{\Ambi}{\mathbb{Q}}
\newcommand{\Flag}{\mathbb{F}}

\ifx\genfrac\sdflkaj\else\fi
\newcommand{\sfrac}[2]{{\textstyle\frac{#1}{#2}}}
\newcommand{\half}{\sfrac{1}{2}}
\newcommand{\ihalf}{\sfrac{i}{2}}

\newcommand{\matr}[2]{\left(\begin{array}{#1}#2\end{array}\right)}

\newcommand{\trans}{{\scriptscriptstyle\mathrm{T}}}

\newcommand{\bigbrk}[1]{\bigl(#1\bigr)}

\newcommand{\bigcomm}[2]{\big[#1,#2\big]}
\newcommand{\comm}[2]{[#1,#2]}

\newcommand{\bigacomm}[2]{\big\{#1,#2\big\}}

\newcommand{\set}[1]{\{#1\}}

\newcommand{\alg}[1]{\mathfrak{#1}}

\newcommand{\nln}{\nonumber\\}

\def\[{\begin{equation}}
\def\]{\end{equation}}
\def\<{\begin{eqnarray}}
\def\>{\end{eqnarray}}
\def\nln{\nonumber\\}

\makeatletter
\let\@keywords\@empty
\let\@subject\@empty
\providecommand{\keywords}[1]{\gdef\@keywords{#1}}
\providecommand{\subject}[1]{\gdef\@subject{#1}}
\def\thetitle{\@title}
\def\theauthor{\@author}
\def\thesubject{\@subject}
\def\thedate{\@date}
\def\thekeywords{\@keywords}
\makeatother
\AtBeginDocument{
\hypersetup{pdftitle={\thetitle}}%
\hypersetup{pdfauthor={\theauthor}}%
\hypersetup{pdfsubject={\thesubject}}%
\hypersetup{pdfkeywords={\thekeywords}}%
}

\makeatletter
\def\mr@ignsp#1 {\ifx\:#1\@empty\else #1\expandafter\mr@ignsp\fi}%
\newcommand{\multiref}[1]{\begingroup
\xdef\mr@no@sparg{\expandafter\mr@ignsp#1 \: }%
\def\mr@comma{}%
\@for\mr@refs:=\mr@no@sparg\do{\mr@comma\def\mr@comma{,}\ref{\mr@refs}}%
\endgroup}
\makeatother

\renewcommand{\eqref}[1]{(\multiref{#1})}

\newcommand{\namedref}[2]{\hyperref[#2]{#1~\ref*{#2}}}

\newcommand{\secref}[1]{\namedref{Sec.}{#1}}

\newcommand{\figref}[1]{\namedref{Fig.}{#1}}

\makeatletter
\newlength{\apb@width}
\newcommand{\autoparbox}[2][c]{\settowidth{\apb@width}{#2}\parbox[#1]{\apb@width}{#2}}
\newcommand{\includegraphicsbox}[2][]{\autoparbox{\includegraphics[#1]{#2}}}
\makeatother

\ifx\href\asklfhas\newcommand{\href}[2]{#2}\fi
\newcommand{\arxivlink}[1]{\href{http://arxiv.org/abs/#1}{arxiv:#1}}

\title{On the Geometry of Null Polygons\texorpdfstring{\\}{ }in Full \texorpdfstring{$\superN=4$}{N=4} Superspace}
\author{Niklas Beisert\texorpdfstring{$^{a,b,d}$}{}, Cristian Vergu\texorpdfstring{$^{a,c,d}$}{}}

\begin{document}

\pdfbookmark[1]{Title Page}{title}

\thispagestyle{empty}
\begin{flushright}\footnotesize
\texttt{\arxivlink{1203.0525}}\\
\texttt{AEI-2012-006}\\%
\texttt{NSF-KITP-12-010}\\%
\end{flushright}
\vspace{1cm}

\begin{center}%
\begingroup\Large\bfseries\thetitle\par\endgroup
\vspace{1cm}%

\begingroup\scshape\theauthor\par\endgroup
\vspace{5mm}%

\begingroup\itshape
$^a$
Institut f\"ur Theoretische Physik\\
Eidgen\"ossische Technische Hochschule Z\"urich\\
Wolfgang-Pauli-Strasse 27, 8093 Z\"urich, Switzerland
\vspace{3mm}

$^b$
Max-Planck-Institut f\"ur Gravitationsphysik\\
Albert-Einstein-Institut\\
Am M\"uhlenberg 1, 14476 Potsdam, Germany
\vspace{3mm}

$^c$
Department of Physics, Brown University\\
Box 1843, Providence, RI 02912, USA
\vspace{3mm}

$^d$
Kavli Institute for Theoretical Physics\\
University of California\\
Santa Barbara, CA 93106, USA
\par\endgroup
\vspace{5mm}

\begingroup\ttfamily
\verb+{nbeisert,verguc}@itp.phys.ethz.ch+
\par\endgroup

\vspace{1cm}

\textbf{Abstract}\vspace{7mm}

\begin{minipage}{12.7cm}
We discuss various formulations of 
null polygons in full, non-chiral $\mathcal{N}=4$ superspace
in terms of spacetime, spinor and twistor variables. 
We also note that null polygons are necessarily 
fat along fermionic directions, 
a curious fact which is compensated by suitable equivalence relations 
in physical theories on this superspace.
\end{minipage}

\end{center}

\newpage

\section{Introduction}
\label{sec:intro}

Recently, light-like Wilson loops in $\mathcal{N} = 4$ super Yang--Mills theory
have become a focus of attention because of their surprising duality to scattering amplitudes
(see \cite{Alday:2008yw} and the special issue \cite{Roiban:2011zz} for reviews).
This duality was inspired by the strong coupling computation of
Alday and Maldacena\ \cite{arXiv:0705.0303} and later understood
as a fermionic T-duality (see\ \cite{arXiv:0807.3196} and also\ \cite{arXiv:0807.3228}).
At weak coupling the duality was confirmed
in refs.~\cite{arXiv:0707.1153, arXiv:0709.2368, arXiv:0712.1223, arXiv:0712.4138, arXiv:0803.1466, arXiv:0803.1465}.
See also ref.~\cite{Bullimore:2011ni} for a proof that the chiral supersymmetric Wilson loop 
yields the same integrand as the scattering amplitudes, as obtained in ref.~\cite{ArkaniHamed:2010kv}.

In the beginning, the duality was between Wilson loops and colour-ordered
MHV scattering amplitudes divided by their tree-level value.
But the scattering amplitudes have a richer structure
and beyond MHV they contain nilpotent invariants when written in superspace.
It was then natural to try to build a modified light-like Wilson loop
which reproduces these nilpotent invariants.
Mason and Skinner constructed such a super Wilson loop in twistor space 
and explicitly worked out its spacetime form to the first two orders in $\theta$ 
\cite{arXiv:1009.2225}
while Caron-Huot constructed a spacetime version in\ \cite{arXiv:1010.1167}.

All of the above constructions for the super Wilson loops either
in spacetime or twistor space have been chiral.
 In a chiral formalism the parity symmetry is not manifest and, for example,
the $\gen{Q}$ and $\gen{\bar{Q}}$ supercharges act in a different way.
In ref.~\cite{CaronHuot:2011ky} Caron-Huot has considered the implications
of a non-chiral formulation.  He found that it is possible to repair
the non-invariance of the remainder function under $\gen{\bar{Q}}$ by adding
a dependence on an antichiral $\bar{\theta}$ Grassmann variable.
The fact that such an expansion in $\bar{\theta}$ is possible had remarkable consequences;
using it, Caron-Huot was able to make a prediction
for the two-loop Grassmann weight-zero part of the super Wilson loop.

This hints that it should be possible to build a super Wilson loop in full superspace.
This belief is reinforced by constructions of
light-like correlation functions
\cite{arXiv:1007.3246, arXiv:1007.3243, arXiv:1009.2488, arXiv:1103.3714, arXiv:1103.4353, arXiv:1108.3557}
which naturally live in full superspace.
However, until now the consequences of this extension to full superspace
have not be worked out in the correlation functions approach.

\medskip

In this paper we set to construct a null polygonal Wilson loop in full superspace.
As we will show below, this is not completely straightforward
since there is no natural notion of straight light-like curves in
superspace which are preserved by superconformal symmetry.
This is in contrast to the bosonic case
where light-like lines are preserved by conformal transformations.
Instead, we realise that we should add eight fermionic directions
to obtain ``fat'' null lines with dimension $1 \vert 8$.
These fat lines are preserved by superconformal transformations.
Importantly, all curves on them are physically equivalent:
All superparticle trajectories are equivalent by means of $\kappa$-symmetry
and likewise Wilson lines due to a flatness constraint of the superspace connection.
Fat lines intersect pairwise in points of full superspace,
which are the vertices of our null polygon.

This spacetime picture can be transformed to ambitwistor space,
which is a non-chiral version of twistor space.
Unfortunately, the ambitwistor theory is poorly understood so
this construction cannot yet be used to directly compute expectation values.
However, we hope that, by comparing to spacetime computations
we will be able to learn how to do perturbation theory in ambitwistor space.
In a companion paper \cite{Beisert:2012xx}
we perform a one-loop computation in spacetime.

\medskip

Most of the above mentioned facts are known from 
various considerations of $\superN=4$ super Yang--Mills theory. 
Here we shall collect and review the geometrical facts which are required 
towards the computation of Wilson loop expectation values 
for null polygons in full $\superN=4$ superspace.
We shall (re)derive them from a purely geometrical perspective, 
and only later connect them to physics.

This paper is organised as follows.
We start in \secref{sec:superspace} by introducing aspects of
$\superN=4$ extended superspace.
We then discuss useful parametrisations
of null polygons in terms of its vertices,
spinor variables and twistor variables
in \secref{sec:polygons}.
A proper definition of the polygon's edges
in terms of fat null lines is the subject of \secref{sec:fat}.
In \secref{sec:curves} we review how to
make physical sense of the segments' fatness.
We conclude in \secref{sec:concl} where we also comment on the duality
between our Wilson loop and scattering amplitudes.

\section{Superspace}
\label{sec:superspace}

We define full (non-chiral) $D=4$, $\superN=4$ superspace
and outline its conformal transformations.

\subsection{Superspace}


Superspace is formulated using spacetime spinors,
therefore let us specify convenient conventions to deal with them
in four dimensions.
All objects will have definite types and positions of spinor indices.
For instance, spacetime coordinates $x$
are represented by a $2\times 2$ hermitian matrix
after multiplying with the 4D Pauli matrices $\sigma$
\[
x^{\beta\dot\alpha}
=\sigma^{\beta\dot\alpha}_\mu x^\mu
=\matr{cc}{t+z&x-iy\\x+iy&t-z}.
\]
Our notation has no implicit rules to move indices to desired places.
Indices can be swapped by transposition ($^\trans$),
or raised and lowered by the Lorentz-invariant antisymmetric matrices
\[
\varepsilon_{\alpha\gamma}=\varepsilon_{\dot\alpha\dot\gamma}
=\varepsilon^{\alpha\gamma}=\varepsilon^{\dot\alpha\dot\gamma}
=\matr{cc}{0&+\\-&0}.
\]
E.g.\ $\varepsilon^2=-1$ will hold for all suitable types of $\varepsilon$.
It is also used to construct the vector products,
for example
\[
x\varepsilon x^\trans
=-x^2\varepsilon,
\qquad
x^\trans\varepsilon x
=-x^2\varepsilon.
\]
Here $x^2$ refers the vector norm which we define as $x^2:=x\cdot x=-t^2+x^2+y^2+z^2$,
i.e.\ the signature of spacetime is ${-}{+}{+}{+}$.


Full non-chiral $\superN=4$ superspace in $D=4$ Minkowski space has a set of $4|16$ real coordinates
\[
X=(x^{\beta \dot\alpha},
\theta^{\beta a},
\bar\theta_b{}^{\dot\alpha}).
\]
We usually do not specify indices, and take $x$ to be a hermitian $2\times 2$ matrix,
while $\theta$ and $\bar\theta$ are hermitian conjugate
$2\times 4$ and $4\times 2$ matrices, respectively
\[\label{eq:reality}
x^\dagger=x,\qquad
\theta^\dagger=\bar\theta,\qquad
\bar\theta^\dagger = \theta.
\]
We follow the convention that in $(3,1)$ Minkowski signature, 
a symbol with bar will denote the complex
conjugate of the same symbol without bar, up to some
simple manipulations. 
All our considerations will be perfectly valid 
in Minkowski signature, although reality conditions 
will not play a significant role.
For most purposes we may work as well with the complexified
superspace where $x,\theta,\bar\theta$ are assumed to be unrelated
complex matrices. Equivalently, in $(2,2)$ split signature,
$x,\theta,\bar\theta$ are unrelated real matrices.
The displayed reality conditions, however, will always refer 
to $(3,1)$ Minkowski signature.

For future use, it makes sense to define the chiral coordinates $x^\pm$
\[\label{eq:chiralx}
x^\pm := x\pm i\theta\bar\theta.
\]
The two pairs of (complex conjugate) coordinates $(x^+,\theta)$ and $(x^-,\bar\theta)$
define chiral and anti-chiral superspace.
They obey the useful identities
\[\label{eq:xpmtheta}
x^++x^-=2x,\qquad
x^+-x^-=2i\theta\bar\theta.
\]
%

\subsection{Conformal Transformations}


Our construction of null lines involves superconformal transformations.
We begin by specifying the translation generators $\gen{P}, \gen{Q}, \gen{\bar Q}$
corresponding to the three coordinates $x,\theta,\bar\theta$ of superspace
\[
\label{eq:susy-gen}
\gen{P}_{\dot\alpha\beta}=\frac{\partial}{\partial x^{\beta\dot\alpha}}\,,
\qquad
\gen{Q}_{a\beta}=
\frac{\partial}{\partial \theta^{\beta a}}
- i \bar\theta_a{}^{\dot\gamma}
\frac{\partial}{\partial x^{\beta\dot\gamma}}\,,
\qquad
\gen{\bar Q}_{\dot\alpha}{}^b=
-\frac{\partial}{\partial \bar\theta_b{}^{\dot\alpha}}
+ i \theta^{\gamma a}
\frac{\partial}{\partial x^{\gamma\dot\alpha}}\,.
\]
For our purposes it will be more convenient to use the language of
variations. Define the variation generator
$\delta:=\Tr (\psi\gen{Q})+\Tr (\gen{\bar Q}\bar\psi)$
with variation parameters $\psi,\bar\psi$.
The corresponding bosonic shift follows by anticommuting two fermionic shifts,
and we can safely disregard it.
The variations of the various superspace coordinates read
\[
\label{eq:translation}
\delta x=-i\psi\bar\theta+i\theta\bar\psi,
\qquad
\delta \theta=\psi,
\qquad
\delta \bar\theta=\bar\psi,
\qquad
\delta x^+=2i\theta\bar\psi,
\qquad
\delta x^-=-2i\psi\bar\theta.
\]


The representation of superconformal boosts
is neither obvious nor simple.
We use a conformal inversion instead,
and derive the boosts from it.
The conformal inversion is most conveniently specified
in terms for the chiral and anti-chiral coordinates
\[
x^\pm \mapsto \varepsilon(x^{\mp\trans})^{-1}\varepsilon,
\qquad
\theta \mapsto -\varepsilon(x^{-\trans})^{-1}\,\bar\theta^\trans M,
\qquad
\bar\theta \mapsto M^{-1}\theta^\trans\, (x^{+\trans})^{-1}\varepsilon.
\]
Here $M$ is some $4\times 4$ symmetric unitary matrix ($M^\trans=M$, $M^\dagger=M^{-1}$)
to specify the action on the fermionic coordinates.
This matrix is necessary for correct transformations under $R$-symmetry.
It is non-canonical since the inversion can be redefined to consist
of the initial inversion operation followed by an $R$-symmetry transformation.  The constraint $M^\trans=M$ is necessary for the inversion transformation to square to the identity.
The inversion of $x$ follows consistently
\[
x\mapsto
\varepsilon(x^{-\trans})^{-1} x^\trans (x^{+\trans})^{-1}\varepsilon.
\]

The representation of boost generators $\gen{K},\gen{S},\gen{\bar S}$
equals translations conjugated by inversions.
The calculation is somewhat lengthy, we merely specify the
final result in the language of variations
\begin{align}
\label{eq:boost}
\delta x&=
-i\theta \bar\rho\varepsilon x^+
-ix^-\varepsilon \rho \bar\theta,
\nln
\delta\theta &=
x^+\varepsilon \rho
-2i\theta  \bar\rho \varepsilon\theta,
\nln
\delta \bar\theta &=
-\bar\rho\varepsilon x^-
-2i\bar\theta \varepsilon \rho \bar\theta,
\nln
\delta x^+&= -2i \theta \bar\rho \varepsilon x^+,
\nln
\delta x^-&=  -2ix^-\varepsilon \rho \bar\theta.
\end{align}
Here, the variation parameters $\rho,\bar\rho$ correspond to
$\gen{\bar S},\gen{S}$, respectively.

\subsection{Null Intervals}
\label{sec:null}

We will be interested in polygons with light-like segments,
so let us discuss intervals
$X_{j,k}=(x_{j,k},\theta_{j,k},\bar\theta_{j,k})$
between two points $X_j=(x_j, \theta_j, \bar{\theta}_j)$ and $X_k=(x_k, \theta_k, \bar{\theta}_k)$ in superspace,
their transformations and the null condition.
In flat bosonic Minkowski space, intervals would simply
be differences of Cartesian coordinates.
However, due to superspace torsion,
the definition of intervals in superspace
includes quadratic terms in the fermionic coordinates in $x_{j,k}$
\[\label{eq:interval}
x_{j,k}:=x_k-x_j-i\theta_k\bar\theta_j+i\theta_j\bar\theta_k,
\qquad
\theta_{j,k}:=\theta_k-\theta_j,
\qquad
\bar\theta_{j,k}:=\bar\theta_k-\bar\theta_j.
\]
The quadratic terms are required to restore exact invariance under
superspace translations \eqref{eq:translation}.
Under superspace boosts \eqref{eq:boost}
the interval transforms as follows
\begin{align}
\label{eq:boostinterval}
\delta x_{j,k}&=
-ix_{j,k}\varepsilon \rho(\bar\theta_j+\bar\theta_k)
-i(\theta_j+\theta_k)\bar\rho\varepsilon x_{j,k}
+\theta_{j,k}(\bar\rho\varepsilon \theta_{j,k}-\bar\theta_{j,k}\varepsilon \rho)\bar\theta_{j,k}
,
\nln
\delta\theta_{j,k}&=
+x_{j,k}\varepsilon \rho
+i\theta_{j,k}(\bar\theta_j+\bar\theta_k)\varepsilon \rho
-i\theta_{j,k} \bar\rho \varepsilon (\theta_j+\theta_{k})
-i(\theta_j+\theta_k) \bar\rho \varepsilon\theta_{j,k},
\nln
\delta\bar\theta_{j,k}&=
-\bar\rho\varepsilon x_{j,k}
+i\bar\rho\varepsilon (\theta_j+\theta_k)\bar\theta_{j,k}
-i\bar\theta_{j,k}\varepsilon \rho (\bar\theta_j+\bar\theta_k)
-i(\bar\theta_j+\bar\theta_k) \varepsilon \rho \bar\theta_{j,k}
.
\end{align}

A suitable definition for null intervals in superspace
consists of the following three conditions 
\[\label{eq:nullinterval}
x_{j,k}^2=0,
\qquad
x_{j,k}^\trans\varepsilon\theta_{j,k}=0,
\qquad
\bar\theta_{j,k}\varepsilon x_{j,k}^\trans=0.
\]
All three of them are required if 
one insists that the null conditions remain stable 
under superconformal transformations:
Translation-invariance \eqref{eq:translation} holds by construction
of the superspace interval.
Invariance under superconformal boosts \eqref{eq:boost}
holds as well, but the confirmation
in terms of \eqref{eq:boostinterval} requires some patience.

The above null conditions imply a host of further relations or formulations.
For instance, \eqref{eq:nullinterval} states that
the spinor indices of
$\theta_{j,k}$ and $\bar\theta_{j,k}$
are collinear with the respective spinor
index of $x_{j,k}$.
This implies the further orthogonality relations
among the fermionic intervals
\[
\theta_{j,k}^\trans\varepsilon\theta_{j,k}=0,
\qquad
\bar\theta_{j,k}\varepsilon \bar\theta_{j,k}^\trans=0.
\]
However, note that the difference of bosonic coordinates $x_k-x_j$
is not exactly null,
but rather $(x_k-x_j)^2=
-\Tr(\theta_{j,k}\bar\theta_j\varepsilon \bar\theta_{j,k}^\trans\theta_j^\trans\varepsilon)$.

Also for the chiral coordinates \eqref{eq:chiralx} there exist
useful definitions of intervals, namely
$(x^+_{j,k},\theta_{j,k})$,
$(x^-_{j,k},\bar\theta_{j,k})$
and the mixed chiral interval
$x^{+-}_{j,k}=-x^{-+}_{k,j}$
with
\begin{align}
\label{eq:mixinter}
x_{j,k}^{+}
&:=
x^+_{k}-x^+_{j},
\nln
x_{j,k}^{-}
&:=
x^-_{k}-x^-_{j},
\nln
x_{j,k}^{+-}
&:=
x^-_{k}-x^+_{j}+2i\theta_j\bar\theta_k
=
x_{j,k}-i\theta_{j,k}\bar\theta_{j,k},
\nln
x_{j,k}^{-+}
&:=
x^+_{k}-x^-_{j}-2i\theta_k\bar\theta_j
=
x_{j,k}+i\theta_{j,k}\bar\theta_{j,k}
.
\end{align}

The null condition can be formulated in terms of chiral and anti-chiral
intervals
\begin{align}
(x^{+}_{j,k})^2&=0,&
x^{+,\trans}_{j,k} \varepsilon \theta_{j,k}&=0,&
\theta_{j,k}^\trans \varepsilon \theta_{j,k}&=0,
\nln
(x^-_{j,k})^2&=0,&
\bar\theta_{j,k}\varepsilon  x^{-,\trans}_{j,k}&=0,&
\bar\theta_{j,k} \varepsilon \bar\theta_{j,k}^\trans&=0,
\nln
(x^{+-}_{j,k})^2 &=0.
\end{align}
%

\section{Null Polygons in Superspace}
\label{sec:polygons}

The definition of null polygons in bosonic Minkowski space
is straight-forward.
The lift to extended superspace is however not so obvious due to torsion.
Here we construct null polygons in superspace and
present three useful parametrisations.

\subsection{Vertices}

\begin{figure}\centering
\includegraphics{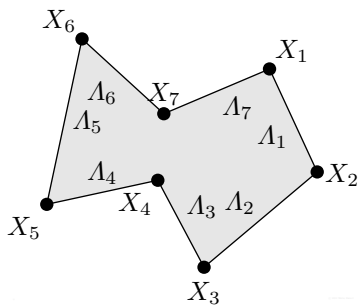}
\caption{Null polygon as a sequence of points in superspace
connected by null line segments.
Indicated are the vertices $X_k$ and spinor variables $\Lambda_j$ 
corresponding to the edges $X_j\to X_{j+1}$.}
\label{fig:Poly}
\end{figure}

A polygon in superspace is specified through
a sequence of vertices $X_k=(x_k,\theta_k,\bar\theta_k)$, $k=1,\ldots,n$,
see \figref{fig:Poly}.
For a null polygon we demand that the segment between
two adjacent vertices is null, cf.\ \secref{sec:null},
\[
x_{k,k+1}^2=0,
\qquad
x_{k,k+1}^\trans\varepsilon\theta_{k,k+1}=0,
\qquad
\bar\theta_{k,k+1}\varepsilon x_{k,k+1}^\trans=0.
\]
The polygon is closed, hence we identify vertex $n+1$ with vertex $1$,
and more generally vertex numbers will be considered modulo $n$.

Let us count the degrees of freedom of the polygon.
Each vertex contributes $4|16$ degrees of freedom.
The null condition for each segment amounts to $1|8$ constraints.
In total, the polygon thus has $3n|8n$ degrees of freedom.

\subsection{Spinor Variables}

For the segment between vertices $k$ and $k+1$ of the polygon,
we solve the null condition in terms of spinor helicity
variables $\Lambda_k:=(\lambda_k, \bar\lambda_k, \eta_k, \bar\eta_k)$,
see \figref{fig:Poly} for the labelling of vertices and edges.
The $\lambda$'s are 2-component bosonic vectors,
the $\eta$'s are 4-component fermionic vectors.
The general solution reads
\[\label{eq:spinor}
x_{k,k+1}=\lambda_k\bar\lambda_k,
\qquad
\theta_{k,k+1}=\lambda_k\eta_k,
\qquad
\bar\theta_{k,k+1}=\bar\eta_k\bar\lambda_k.
\]
%
Compatibility with the reality condition \eqref{eq:reality}
implies the following complex conjugation properties
\[\label{eq:spinorreality}
\lambda^\dagger_k = \pm \bar\lambda_k,
\qquad
\eta^\dagger_k = \pm \bar\eta_k,
\]
with a common sign for both relations.
The above parametrisation is invariant under the rescaling
(reality conditions imply that $z$ is a pure complex phase)
\[\label{eq:spinorgauge}
\lambda_k\mapsto z_k\lambda_k,
\qquad
\bar\lambda_k\mapsto z^{-1}_k\bar\lambda_k,
\qquad
\eta_k\mapsto z^{-1}_k\eta_k,
\qquad
\bar\eta_k\mapsto z_k\bar\eta_k.
\]
Thus, we have $3|8$ degrees of freedom for each segment,
but $4|16$ constraints for the closure of the polygon.
In total there are $(3n-4)|(8n-16)$ degrees of freedom
for the spinor variables.
As the spinor variables are invariant
under translations,
a reference vertex provides the remaining $4|16$ degrees of freedom
for the polygon.

Let us next derive the superconformal transformations
of the spinor variables.
As the intervals are translation-invariant,
so are the spinor variables.
For the superconformal boosts, we substitute the
definition \eqref{eq:spinor}
into the boost transformation of the interval
\eqref{eq:boostinterval}
\begin{align}
\delta(\lambda_k\bar\lambda_k)=&\
\lambda_k\bigbrk{
-i(1-i\eta_k\bar\eta_k)(\bar\lambda_k\varepsilon \rho \bar\eta_k)\bar\lambda_k
-2i\bar\lambda_k\varepsilon \rho \bar\theta_{k}
}
\nln&
+\bigbrk{
-i(1+i\eta_k\bar\eta_k)(\eta_k \bar\rho\varepsilon \lambda_k)\lambda_k
-2i\theta_k \bar\rho\varepsilon \lambda_k
}\bar\lambda_k,
\nln
\delta(\lambda_k\eta_k)=&\
\lambda_k\bigbrk{(1+i\eta_k\bar\eta_k)\bar\lambda_k+2i\eta_k\bar\theta_k}\varepsilon \rho
\nln&
-2i\lambda_k\eta_k  \bar\rho \varepsilon \theta_k
-2i\lambda_k\eta_k  \bar\rho \varepsilon \lambda_k\eta_k
-2i\theta_k \bar\rho \varepsilon \lambda_k\eta_k,
\nln
\delta(\bar\eta_k\bar\lambda_k)=&
-2i\bar\eta_k\bar\lambda_k \varepsilon \rho \bar\theta_k
-2i\bar\theta_k \varepsilon \rho \bar\eta_k\bar\lambda_k
-2i\bar\eta_k\bar\lambda_k \varepsilon \rho \bar\eta_k\bar\lambda_k
\nln&
+\bar\rho\varepsilon\bigbrk{
-(1-i\eta_k\bar\eta_k)\lambda_k
+2i\theta_k\bar\eta_k}
\bar\lambda_k.
\end{align}
These transformations can be split up into boost transformations
for the spinor variables essentially because the null condition
is superconformally invariant
\begin{align}
\label{eq:spinorboost}
\delta \lambda_k=&
+i\Tr(\rho\bar\alpha_k-\alpha_k\bar\rho)\lambda_k
-i(1+i\eta_k\bar\eta_k)(\eta_k \bar\rho\varepsilon \lambda_k)\lambda_k
-2i\theta_k \bar\rho\varepsilon \lambda_k,
\nln
\delta\bar\lambda_k=&
-i\Tr(\rho\bar\alpha_k-\alpha_k\bar\rho)\bar\lambda_k
-i(1-i\eta_k\bar\eta_k)(\bar\lambda_k\varepsilon \rho \bar\eta_k)\bar\lambda_k
-2i\bar\lambda_k\varepsilon \rho \bar\theta_k,
\nln
\delta\eta_k=&
-i\Tr(\rho\bar\alpha_k-\alpha_k\bar\rho)\eta_k
+(1+i\eta_k\bar\eta_k)\bar\lambda_k\varepsilon \rho
+2i\eta_k\bar\theta_k\varepsilon \rho
\nln&
-i(1-i\eta_k\bar\eta_k)(\eta_k \bar\rho\varepsilon \lambda_k)\eta_k
-2i\eta_k  \bar\rho \varepsilon \theta_k,
\nln
\delta\bar\eta_k
=&
+i\Tr(\rho\bar\alpha_k-\alpha_k\bar\rho)\bar\eta_k
-i(1+i\eta_k\bar\eta_k)(\bar\lambda_k\varepsilon \rho \bar\eta_k)\bar\eta_k
-2i\bar\theta_k \varepsilon \rho \bar\eta_k
\nln&
-(1-i\eta_k\bar\eta_k)\bar\rho\varepsilon\lambda_k
+2i\bar\rho\varepsilon\theta_k\bar\eta_k.
\end{align}
Here the $\alpha$'s parametrise the transformation of
the unphysical degree of freedom in \eqref{eq:spinorgauge}.

\subsection{Twistor Variables}

The above boost transformations of the spinor variables
\eqref{eq:spinorboost} are somewhat intransparent.
It is convenient to introduce so-called momentum twistor variables
\cite{arXiv:0905.1473,Mason:2009qx} (cf.\ reviews in \cite{Bullimore:2010pj,Adamo:2011pv})
to parametrise our null polygon. They will turn out to transform nicely.
A momentum twistor $W_k$ and its conjugate $\bar W_k$ are
complex projective $4|4$ vectors defined by
\begin{align}
W_k&:=(-\ihalf\lambda_k^\trans\varepsilon,\mu_k,\chi_k),
&
\mu_k&:=\lambda_k^\trans\varepsilon x^+_{k},
&
\chi_k&:=\lambda_k^\trans\varepsilon \theta_k,
\nln
\bar W_k&:=(\bar\mu_k,-\ihalf\varepsilon\bar \lambda_k^\trans,\bar\chi_k),
&
\bar\mu_k&:=-x^-_{k} \varepsilon \bar\lambda_k^\trans,
&
\bar\chi_k&:=-\bar\theta_k\varepsilon \bar\lambda_k^\trans.
\end{align}
Reality conditions for the twistors follow from
\eqref{eq:reality,eq:spinorreality}. 
They impose the hermitian signature $(2,2|4)$
on $(W_k,\bar W_k)$ 
by means of a conjugation matrix $C$ written in $2,2,4$ block form%
\footnote{The sign in the reality condition 
specifies an orientation of the corresponding polygon segment.
The conjugation property can be fixed to
$W_k^\dagger = C \bar W_k$ by rescaling the definition of $\bar W$ by $\pm 1$.}
\[\label{eq:twistorreality}
W_k^\dagger = \pm C \bar W_k,
\qquad
C=\matr{cc|c}{0&1&0\\1&0&0\\\hline0&0&1}.
\]

As before, the superconformal transformations follow
by substituting the definitions.
For translations we obtain
from \eqref{eq:translation} simply
\begin{align}
\delta\lambda_k^\trans&=0,
&
\delta\bar\lambda_k^\trans&=0,
\nln
\delta\chi_k&=\lambda_k^\trans\varepsilon \psi,
&
\delta\bar\chi_k&=-\bar\psi\varepsilon \bar\lambda_k^\trans,
\nln
\delta\mu_k&=2i\chi_k\bar\psi,
&
\delta\bar\mu_k&=-2i\psi\bar\chi_k.
\end{align}
Boosts follow from \eqref{eq:boost,eq:spinorboost}
\begin{align}
\delta\lambda_k^\trans&=\beta_k\lambda_k^\trans
+2i \chi_k  \bar\rho,
&
\delta\bar\lambda_k^\trans&=\bar\beta_k\bar\lambda_k^\trans
-2i \rho \bar\chi_k ,
\nln
\delta\chi_k&=\beta_k\chi_k
+\mu_k\varepsilon \rho,
&
\delta\bar\chi_k&=\bar\beta_k\bar\chi_k
-\bar\rho \varepsilon \bar\mu_k,
\nln
\delta\mu_k&=\beta_k\mu_k,
&
\delta\bar\mu_k&=\bar\beta_k\bar\mu_k.
\end{align}
The $\beta$'s correspond to rescalings of the twistors $W_k$ and $\bar W_k$.
Due to the projective nature of twistors, the $\beta$'s are inessential,
we can nevertheless state their expression in terms of spinor variables
\begin{align}
\beta_k&:=+i\Tr(\rho\bar\alpha_k-\alpha_k\bar\rho-2\varepsilon\theta_k \bar\rho)
-i(1+i\eta_k\bar\eta_k)(\eta_k \bar\rho\varepsilon \lambda_k),
\nln
\bar\beta_k&:=-i\Tr(\rho\bar\alpha_k-\alpha_k\bar\rho+2\rho \bar\theta_k  \varepsilon)
-i(1-i\eta_k\bar\eta_k)(\bar\lambda_k \varepsilon \rho \bar\eta_k  ).
\end{align}
In summary, the twistors $W_k$ and $\bar W_k$ transform as projective fundamental
and anti-fundamental representations of the superconformal algebra $\alg{psu}(2,2|4)$.

It is now straight-forward to construct the projective invariants
\begin{align}
W_j \bar W_k
&=-\ihalf\lambda_j^\trans\varepsilon\bar\mu_k-\ihalf \mu_j\varepsilon\bar \lambda_k^\trans+\chi_j\bar\chi_k
\nln
&=\ihalf\lambda_j^\trans \varepsilon
(x^-_k - x^+_j+ 2i\theta_j   \bar\theta_k)
\varepsilon \bar\lambda_k^\trans
=\ihalf\lambda_j^\trans \varepsilon x^{+-}_{j,k} \varepsilon \bar\lambda_k^\trans.
\end{align}
They transform as $\delta (W_j \bar W_k)=(\beta_j+\bar\beta_k)W_j \bar W_k$.
Proper invariants can be obtained as functions of these with
vanishing weights in each of the twistors variables $W_k$ and, separately, 
their conjugates $\bar W_k$.

Note that these momentum twistor variables are constrained.
By virtue of \eqref{eq:xpmtheta} one finds
\[
W_k\bar W_k=
-\ihalf\lambda_k^\trans \varepsilon
(x^+_k-x^-_k - 2i\theta_k\bar\theta_k)
\varepsilon \bar\lambda_k^\trans
=0.
\]
This means that the pair $W_k,\bar W_k$ actually defines
a (real) ambitwistor.
Likewise one finds that contractions of adjacent twistors vanish
\begin{align}
W_k\bar W_{k+1}&=
+\ihalf\lambda_k^\trans \varepsilon \lambda_k
(1-i\eta_k\bar\eta_k)
\bar\lambda_k
\varepsilon \bar\lambda_{k+1}^\trans
=0,
\nln
W_{k+1}\bar W_k&=
-\ihalf\lambda_{k+1}^\trans \varepsilon\lambda_k
(1+i\eta_k\bar\eta_k)
\bar\lambda_k\varepsilon \bar\lambda_k^\trans
=0.
\end{align}
We shall refer to a sequence $W_k,\bar W_k$, $k=1,\ldots,n$,
subject to the constraints
\[\label{eq:momentumambi}
W_j\bar W_k=0\quad \text{for}\quad |j-k|\leq 1
\]
as momentum ambitwistors \cite{arXiv:0905.1473,Mason:2009qx}.

We can now count the real degrees of freedom of the twistor variables.
Both $W_k$ and $\bar W_k$ contribute $4|4$ degrees of freedom.
Independent rescalings of $W_k$ and $\bar W_k$ eliminate
two degrees of freedom, and the ambitwistor condition a third one.
Each ambitwistor thus has $5|8$ degrees of freedom.
There are two additional constraints for each pair of adjacent
vertices, leaving $3n|8n$ degrees of freedom.
This matches precisely the previous counting for the null polygon.
It shows that a null polygon in superspace is described
by a sequence a momentum ambitwistors.

\subsection{Comparison}

We have discussed three different formulations for null polygons in superspace:
\begin{itemize}
\item
The first one specifies the vertices $X_k=(x_k,\theta_k,\bar\theta_k)$. Two adjacent vertices are constrained
to be null-separated.

\item
The second formulation specifies the segments in terms of spinor variables
$\Lambda_k=(\lambda_k,\bar\lambda_k,\eta_k,\bar\eta_k)$.
The null conditions are automatically satisfied, but
constraints are needed to guarantee closure of the polygon.
This formulation is invariant under translations,
a reference vertex is needed to locate the polygon in superspace.

\item
A final description uses momentum ambitwistors $(W_k,\bar W_K)$
to describe the segments and vertices.
Three constraints per segment are needed
to guarantee that the segments intersect properly.

\end{itemize}
In all cases, the polygon is described by $3n|8n$ degrees of freedom,
and we displayed their relations explicitly.

Let us compare this to the case of null polygons in chiral superspace
which has $4|8$ coordinates only
(the anti-chiral case is equivalent).
The above discussion fully applies
through projection of the full superspace $(x,\theta,\bar\theta)$
onto the chiral subspace $(x^+,\theta)$;
in effect, one disregards all $\bar\theta$'s, $\bar\eta$'s and $\bar W$'s.
The chiral null polygon is then described by $3n|4n$ degrees of freedom.
There is, however, one noteworthy difference:
When discarding the $\bar W$'s,
all constraints on chiral momentum twistors drop out.
Unconstrained chiral momentum twistors provide all
the necessary $3n|4n$ degrees of freedom of the polygon!
This crucial benefit comes along with the minor shortcoming
that chiral superspace requires either $(2,2)$ split signature
or complexified Minkowski space.
If reality conditions for $(3,1)$ signature
are imposed on chiral momentum twistors,
one indeed recovers the conjugate twistors
along with the constraints.

Finally, we compare these two cases to the purely bosonic case
by disregarding all fermionic components. The bosonic null polygon is
described by $3n$ degrees of freedom. The formulation in terms of
momentum twistors is equivalent to the formulation in terms of
momentum ambitwistors.
The two are related by the identification
$\bar W_{k,A} = \varepsilon_{ABCD}W_{k-1}^BW_k^CW_{k+1}^D$
up to an inessential factor.
It automatically implies the momentum ambitwistor constraints \eqref{eq:momentumambi}.
Unfortunately, in the supersymmetric case,
the tensor $\varepsilon_{ABCD}$ is not invariant, and
a supersymmetrisation does not exist. Hence,
we are forced to use the ambitwistor formulation
for the full superspace.

\section{Fat Null Polygons}
\label{sec:fat}

Next we wish to define the null polygon curve.
Here we encounter an interesting surprise.

\subsection{Thin Segments}

\begin{figure}
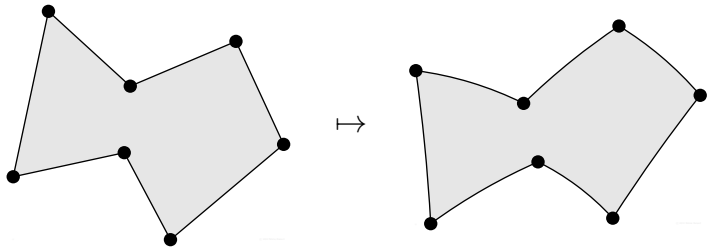
\centering
\includegraphicsbox{FigPolyMap1.mps}
\quad
$\mapsto$
\quad
\includegraphicsbox{FigPolyMap2.mps}
\caption{Conformal transformations in superspace
map straight null line segments to curved ones.}
\label{fig:PolyMap}
\end{figure}

So far we have merely defined the vertices. Two adjacent vertices $k$ and $k+1$
are null-separated, and we shall connect them by a null curve.
The obvious choice is
\begin{align}
\label{eq:linearsegment}
x(\tau)&=x_k+(\lambda_k\bar\lambda_k+i\lambda_k\eta_k\bar\theta_k-i\theta_k\bar\eta_k\bar\lambda_k)\tau,
\nln
\theta(\tau)&=\theta_k+\lambda_k\eta_k\tau,
\nln
\bar\theta(\tau)&=\bar\theta_k+\bar\eta_k\bar\lambda_k\tau.
\end{align}
Unfortunately, it turns out that this kind of curve is not stable under 
a superconformal boost transformation:
In the above curve all coordinates are linear in $\tau$.
After the transformation, the coordinates are not linear.
In the bosonic case, a compensating reparametrisation
$\tau\to\tau'$ is required to recover linearity.
In the extended supersymmetric case, such a reparametrisation
does not exist in general.
To see this, let us consider $\theta(\tau)$.
The second derivative $\ddot\theta$ originally vanishes.
For the boost \eqref{eq:boost} of the curve we find
\[
\frac{d^2 \delta\theta}{d\tau^2} =
2i(\eta_k\bar\eta_k)\lambda_k\bar\lambda_k\varepsilon\rho
-4i(\eta_k\bar\rho \varepsilon\lambda_k)\lambda_k\eta_k
\stackrel{?}{=}\frac{d^2 \delta \tau}{d\tau^2}\, \dot\theta.
\]
The identity on the right hand side is the condition
for linearity up to reparametrisation of $\tau$.
The second term in the middle is indeed of the desired form
with $d^2 \delta \tau/d\tau^2=-4i(\eta_k\bar\rho \varepsilon\lambda_k)$
because $\dot\theta=\lambda_k\eta_k$.
The first term in the middle, however, is not. It would require $\rho$ to be collinear to $\eta_k$
which generically does not hold, certainly not for all polygon segments.
In conclusion, boost transformations map polygons constructed
from naive straight null segments \eqref{eq:linearsegment}
to some other shape, cf.\ \figref{fig:PolyMap}.
Furthermore, we did not find a suitable alternative
definition for straight null curves which has this stability property.
This may seem unfortunate because Wilson loops on such null polygons
would appear not to transform nicely, and we could not make use of
superconformal symmetry.
As we shall see shortly, this in fact does not pose a problem.

\subsection{Fat Null Lines}

There is an alternative characterisation of straight null lines
in bosonic spacetime which we can use for superspace as well:
Consider two fixed points $x_0$ and $x_1$
which are null-separated.
A straight null line passing through $x_0$ and $x_1$
is the set of all points $x$ which are null-separated from both $x_0$ and $x_1$.
This defines a straight line because any three null vectors
$x-x_0$, $x_1-x$ and $x_0-x_1$ in Minkowski space
which add up to zero are necessarily collinear. 
This definition is manifestly conformal because the null condition is.
Moreover it carries over to superspace straight-forwardly.

\begin{figure}\centering
\includegraphics{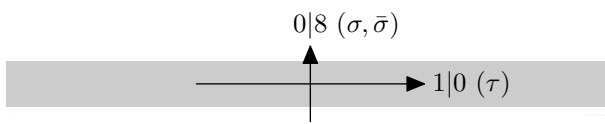}
\caption{A fat null line parametrised through
one bosonic coordinate $\tau$ and 8 fermionic coordinates
$\sigma,\bar\sigma$.}
\label{fig:FatLine}
\end{figure}

Consider therefore two null-separated points $X_0$ and $X_1$ in superspace.
According to \eqref{eq:spinor} we can write the superspace interval
\eqref{eq:interval} as
$x_{0,1}=\lambda\bar\lambda$,
$\theta_{0,1}=\lambda\eta$,
$\bar\theta_{0,1}=\bar\eta\bar\lambda$.
This provides us with a parametrisation of $X_1$ in terms
of $X_0$ and the spinors $\lambda,\bar\lambda,\eta,\bar\eta$
\[
x_1=x_0+\lambda\bar\lambda+i\lambda\eta\bar\theta_0-i\theta_0\bar\eta\bar\lambda,
\qquad
\theta_1=\theta_0+\lambda\eta,
\qquad
\bar\theta_1=\bar\theta_0+\bar\eta\bar\lambda.
\]
All points $X$ at null-separation to $X_0$ must therefore be of the same form but
with different $\lambda',\bar\lambda',\eta',\bar\eta'$.
Null-separation from $X_1$ then merely forces $\lambda'\sim\lambda$ and
$\bar\lambda'\sim\bar\lambda$. Hence we can write the most general solution
as
\[\label{eq:fatcoords}
x=
x_{0}
+\tau\lambda\bar\lambda
+i\lambda\sigma\bar\theta_0
-i\theta_0\bar\sigma\bar\lambda,
\qquad
\theta=\theta_0+\lambda\sigma,
\qquad
\bar\theta=\bar\theta_0+\bar\sigma\bar\lambda.
\]
The solution $X(\tau,\sigma,\bar\sigma)$ is parametrised explicitly
through one bosonic coordinate $\tau$ and a pair of complex conjugate 4-component
fermionic coordinates $(\sigma,\bar\sigma)$.
Curiously, the null line in superspace is ``fattened'' by $8$ fermionic coordinates,
see \figref{fig:FatLine}, cf.\ \cite{Witten:1978xx,Isenberg:1978kk,Harnad:1988rs}.%
\footnote{The fattening (by 4 fermionic coordinates) also applies to null polygons in chiral
superspace.}

The fatness of the null line explains our difficulty
in finding a proper straight line between
two null-separated vertices.
With regard to superconformal transformations,
a fat null line is a very natural object,
its shape manifestly remains stable.
Conversely, there appears to be no distinguished submanifold
of dimension $1|0$.
Our attempt \eqref{eq:linearsegment}
to set $\sigma=\eta\tau$ and $\bar\sigma=\bar\eta\tau$
is one possibility,
but there is nothing that prevents conformal transformations
from distorting our choice.
In \secref{sec:curves} we shall explain that all
curves on a fat null line are physically equivalent.
In other words, a fat null line actually defines
a physically unique curve.

\subsection{Ambitwistors}

Before we continue with the physical implication of fat lines,
let us return to the insight that null polygons are specified by
a sequence of ambitwistors, and let us take it seriously
(see \cite{Bullimore:2010pj,Adamo:2011pv} for reviews of twistors
and ref.~\cite{Manin:1988ds} for an in-depth discussion of the relevant twistor space geometry).

A twistor $W=(-\ihalf \lambda^\trans\varepsilon,\mu,\chi)$
describes a null subspace of superspace
through the equations
for the chiral coordinates $(x^+,\theta)$
\[\label{eq:twistor}
\lambda^\trans\varepsilon x^+=\mu,
\qquad
\lambda^\trans\varepsilon \theta=\chi.
\]
These $2|4$ equations constrain as many coordinates of (complexified) superspace.
Embedding the twistor into chiral superspace, the dimension is $2|4$.
We can parametrise the solution explicitly
through a 2-component bosonic vector $\bar\kappa$ and a 4-component
fermionic vector $\sigma$
\[\label{eq:twistorsol}
x^+(\bar\kappa,\sigma) = x^+_0 + \lambda\bar\kappa,
\qquad
\theta(\bar\kappa,\sigma) = \theta_0 + \lambda\sigma.
\]
Here $x^+_0,\theta_0$ are particular solutions of the inhomogeneous equations.
In full superspace, the anti-chiral coordinates $\bar\theta$ are unconstrained,
and hence the dimension of the twistor in full superspace is $2|12$.

\begin{figure}\centering
\includegraphics{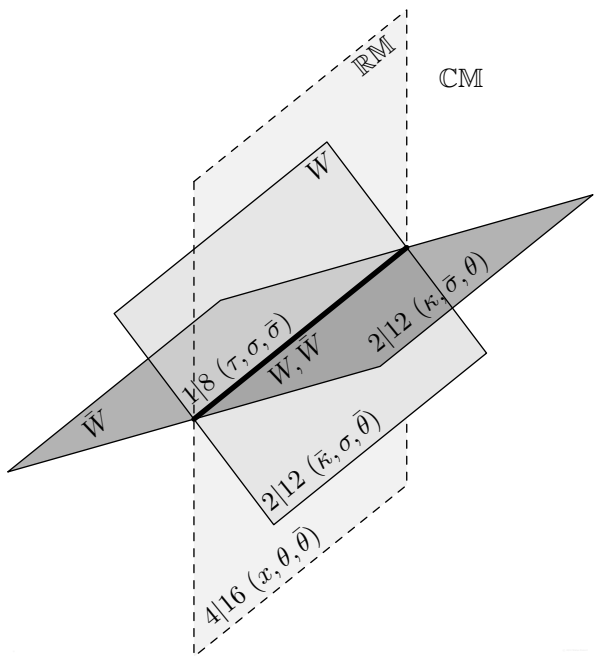}
\caption{A fat null line as the intersection of two complex conjugate twistors
$W,\bar W$. The twistor subspaces reside in complexified superspace $\Complex\Mink$
whereas their intersection is contained in real superspace $\Reals\Mink$.}
\label{fig:TwistorIntersect}
\end{figure}

A conjugate twistor $\bar W$ describes an analogous subspace
\[\label{eq:twistorconj}
-x^-\varepsilon\bar\lambda^\trans =\bar\mu,
\qquad
-\bar\theta\varepsilon\bar\lambda^\trans =\bar\chi.
\]
Superficially, the intersection of the subspaces given by $W$ and $\bar W$
is a space of codimension $4|8$, i.e.\ of dimension $0|8$.
This simple consideration misses the fact that the two twistor equations
are generally incompatible because of the relation \eqref{eq:xpmtheta} between $x^+$ and $x^-$.
Compatibility requires the ambitwistor condition $W\bar W=0$:
\[
0=
\lambda^\trans\varepsilon(x^+-x^--2i\theta\bar\theta)\varepsilon\bar\lambda^\trans
=
\mu \varepsilon\bar\lambda^\trans
+\lambda^\trans\varepsilon \bar\mu
+2i\chi\bar\chi
=2i W\bar W.
\]
The resulting intersection is thus bigger by one bosonic dimension,
namely it has dimension $1|8$,
see \figref{fig:TwistorIntersect} for an illustration
of the twistors and their intersection.
Note that the intersection is contained in real superspace.
It is given precisely by the above explicit parametrisation of the
fat null line in \eqref{eq:fatcoords}.
Note that the chiral coordinates $x^\pm$ both take the
predicted form \eqref{eq:twistorsol} for chiral twistors
for a suitable choice of $\bar\kappa,\kappa$
\[
x^+=x^+_0+\lambda(\tau\bar\lambda+i\sigma\bar\sigma\bar\lambda+2i\sigma\bar\theta_0),
\qquad
x^-=x^-_0+(\tau\lambda-i\sigma\bar\sigma \lambda-2i\theta_0\bar\sigma)\bar\lambda.
\]
Bosonically, an ambitwistor describes a null line.
In superspace, however, the null line is fattened by 
8 real fermionic coordinates, see \figref{fig:FatLine}.
Under superconformal transformations the
ambitwistor $(W,\bar W)$ transforms as a complex conjugate pair of projective fundamental
representations. The corresponding fat null line transforms accordingly.

\begin{figure}\centering
\includegraphics{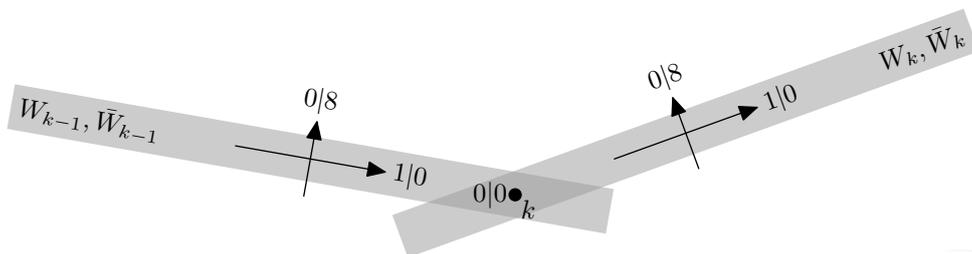}
\caption{Two fat null lines intersect in a (thin) point of dimension $0|0$.}
\label{fig:LineIntersect}
\end{figure}

\begin{figure}\centering
\includegraphics{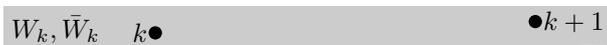}
\caption{Two null-separated points specify a unique twistor.}
\label{fig:ConnectingTwistor}
\end{figure}

We have seen above that a null polygon in superspace
can be given in terms of a sequence of ambitwistors $(W_k,\bar W_k)$.
Taken at face value, our polygon can be viewed as a sequence of
fat null lines. The additional conditions
\[
W_{k-1}\bar W_k=W_k\bar W_{k-1}=0
\]
ensure that two consecutive fat segments intersect.
Although they are fat, they generically intersect in a (thin) point of dimension $0|0$,
namely the vertex $X_k=(x_k,\theta_k,\bar\theta_k)$, see \figref{fig:LineIntersect}.
This is how the vertices are specified by a sequence of ambitwistors.
Let us also remark that there is a unique ambitwistor which connects
two null-separated points, see \figref{fig:ConnectingTwistor}.
This is how the ambitwistors are specified by a sequence of vertices.

\subsection{Dual Polygon in Ambitwistor Space}

A null polygon consists of a sequence of vertices and edges.
The vertices are points $X_k=(x_k,\theta_k,\bar\theta_k)$
in $\superN=4$ Minkowski superspace $\Mink=\Reals^{3,1}\times\Complex^{0|8}=\Reals^{3,1|16}$.
As described above, the edges are fat null lines in $\Mink$.
Alternatively, the edges can be specified through
a sequence of ambitwistors $(W_k,\bar W_k)$.
Now we can also view an ambitwistor as a point in ambitwistor space $\Ambi$.
When the latter points are connected by edges,
we obtain a dual polygon in ambitwistor space \cite{Bullimore:2010pj}.
Let us briefly discuss the nature of this dual polygon.

We specify an ambitwistor $(W,\bar W)$
through a twistor $W\in \Complex^{2,2|4}\backslash\set{0}$
and its complex conjugate $\bar W$,
which is hence not an independent quantity.%
\footnote{Very often in discussions of twistor space, 
the corresponding Minkowski space is assumed to have
complex or $(2,2)$ split signature. For our purposes
there is no need to deviate from real $(3,1)$ signature in what follows.
To translate the discussion to complex signature one would 
complexify real spaces and double complex spaces, 
e.g.\ $\Reals^{3,1}\mapsto \Complex^{4}$
and $\Complex\Projective^{1,2}\mapsto \Complex\Projective^{3}\times\Complex\Projective^{3}$.
To translate to split signature instead, 
one chooses a different real form for the complexified spaces,
e.g.\ $\Reals^{3,1}\mapsto \Reals^{2,2}$
and $\Complex\Projective^{1,2}\mapsto \Reals\Projective^{3}\times\Reals\Projective^{3}$.}
Ambitwistors are projectively identified, i.e.\
$(W,\bar W)\simeq (zW,\bar z\bar W)$ for any $z\in\Complex^{*}$.
Moreover, they satisfy the condition $W\bar W=0$.
Altogether this defines a $5|8$-dimensional real subspace $\Ambi$ 
of the complex projective identification of $\Complex^{2,2|4}\backslash\set{0}$,
see \figref{fig:Ambi}.
The space $\Ambi$ will be called (real) ambitwistor space 
(in the twistor space literature it is usually called 
the space of projective null twistors $\mathbb{PN}$).

\begin{figure}\centering
\includegraphics{FigTwistorSpace.mps}
\caption{
The projective identification of points in $\Complex^{2,2|4}\backslash\set{0}$
splits into three components
$\Complex\Projective^{1,2|4}$ ($W\bar W>0$),
$\Complex\Projective^{2,1|4}$ ($W\bar W<0$)
and ambitwistor space $\Ambi$ ($W\bar W=0$).
The latter (conical surface) has real dimension $5|8$.}
\label{fig:Ambi}
\end{figure}

Consider now the situation at a vertex of the polygon in $\Mink$.
It is described by two fat null lines which meet in a point.
They correspond to two ambitwistors $(W,\bar W)$ and $(W',\bar W')$
which obey the additional condition $W\bar W'=W'\bar W=0$
that makes the associated lines intersect.
The latter condition implies that all the points
on the $\Complex\Projective^1$
joining $(W,\bar W)$ and $(W',\bar W')$
\[
(zW+z'W',\bar z\bar W+\bar z'\bar W')
\quad \mbox{for all }z,z'\in\Complex
\]
are also ambitwistors because they satisfy
\[(zW+z'W')(\bar z \bar W+\bar z'\bar W')=
z\bar z W\bar W
+z\bar z'W\bar W'
+z'\bar z W'\bar W
+z'\bar z' W'\bar W'
=0
.\]
In other words,
the points $W$ and $W'$ are connected by a $\Complex\Projective^1$
which resides entirely within ambitwistor space $\Ambi$.%
\footnote{We thank David Skinner for pointing out this interpretation.}
Hence, the dual of two intersecting lines in $\Mink$
are two points in $\Ambi$ joined by a $\Complex\Projective^1$ inside $\Ambi$.

\begin{figure}\centering
\includegraphicsbox{FigPolySpacetime.mps}
\quad$\longleftrightarrow$\quad
\includegraphicsbox{FigPolyTwistor.mps}
\caption{A fat null polygon in $\Mink$ composed from $\Reals^{1|8}$'s
and the dual fat polygon in $\Ambi$ composed from $\Complex\Projective^1$'s.}
\label{fig:TwistorDual}
\end{figure}

We conclude that the dual of a null polygon in $\Mink$ is a
polygon in $\Ambi$ whose edges are $\Complex\Projective^1$'s,
see \figref{fig:TwistorDual}.
Incidentally the edges of the dual polygon
are $2|0$-dimensional, i.e.\ they are also fat,
moreover along bosonic directions.
In fact, this duality is one-to-one
because a $\Complex\Projective^1$ in $\Ambi$
also describes precisely a single point in $\Mink$:
A $\Complex\Projective^1$ can be specified by two points
$W,W'\in \Complex^{2,2|4}\backslash\set{0}$
which amounts to $16|16$ real degrees of freedom.
They must satisfy the 4 real constraints
$W\bar W=W\bar W'=W'\bar W=W'\bar W'=0$.
Furthermore, any pair of complex linear combinations
of $W$ and $W'$ describes the same $\Complex\Projective^1$
which removes another 8 real degrees of freedom.
Hence, the embedding of a $\Complex\Projective^1$ into $\Ambi$
has $4|16$ moduli which represents a point in $\Mink$.
Geometrically, the $\Complex\Projective^1$
is the sphere which describes
the set of all null directions around a point.

Finally, let us daydream about a combination $\Mink\Ambi$ of
Minkowski space $\Mink$ and ambitwistor space $\Ambi$
which may have some use.
The points of this space describe points in $\Mink$ along
with a null line that passes through the point.
Alternatively, it is a point in $\Ambi$ along with a
$\Complex\Projective^1$ in $\Ambi$ that passes through the point.
Both of these interpretations lead to a dimension
of $6|16=(4|16)+(2|0)=(5|8)+(1|8)$.
The space $\Mink\Ambi$ can be called the space of null rays in $\Mink$,
i.e.\ points together with a null direction.
Technically, a point in $\Mink\Ambi$ is given by a
point $X=(x,\theta,\bar\theta)\in \Mink$
and an ambitwistor $(W,\bar W)\in\Ambi$
subject to the conditions specified in \eqref{eq:twistor,eq:twistorconj}.

A null polygon can be mapped to this space as a polygon with twice as many
vertices and edges. The vertices in $\Mink\Ambi$ correspond to the rays at the
beginning and end of each of the edges. The edges connect the points
along fibres of $\Mink$ and $\Ambi$ in an alternating fashion, see
\figref{fig:Double}. The nice feature of this representation is
that it includes both the spacetime polygon and the twistor polygon
as projections onto the spaces $\Mink$ and $\Ambi$, respectively.

\begin{figure}\centering
\includegraphics{FigPolyDouble.mps}
\caption{A null polygon in the combined space $\Mink\Ambi$.
Horizontal segments are fat null lines in $\Mink$ and points in $\Ambi$.
Conversely, vertical segments are points in $\Mink$ and $\Complex\Projective^1$'s in $\Ambi$.}
\label{fig:Double}
\end{figure}

Finally, we can note that the complexifications 
of the above spaces have representations as various 
flag manifolds of $\Complex^{4|4}$, see e.g.\ \cite{Manin:1988ds,Howe:1995md}.%
\footnote{We thank David Mesterhazy and David Skinner for discussions.}
Chiral twistor space $\Complex\Projective^{3|4}$ equals
the flag manifold $\Flag_{1|0}$ 
while anti-chiral twistor space equals the dual flag manifold $\Flag_{3|4}$.
Chiral superspace corresponds to $\Flag_{2|0}$ while 
antichiral superspace is the dual $\Flag_{2|4}$.
Combinations of these flags yield the above spaces in an obvious fashion:
Ambitwistor space is a combination of
the two chiral twistor spaces $\Ambi=\Flag_{1|0;3|4}$. 
Full superspace is a combination of the two chiral superspaces
$\Mink=\Flag_{2|0;2|4}$. 
The space of null rays is $\Mink\Ambi=\Flag_{1|0;2|0;2|4;3|4}$.
The latter three spaces are self-dual and they have real slices
corresponding to Minkowski signature.

\section{Curves on Fat Null Lines}
\label{sec:curves}

In this section we will review the physical equivalence
of all curves on a fat null line
for the cases of the trajectory of the $\superN=4$ supersymmetric particle
and for Wilson lines in $\superN=4$ supersymmetric Yang--Mills theory.

\subsection{The Superparticle and \texorpdfstring{$\kappa$}{kappa}-Symmetry}
\label{sec:kappa-sym}

Physically, we can think of a Wilson loop as the phase picked up by a
non-dynamical charged particle moving in its own gauge field.
 In the case of super-Wilson loops, the same holds but this time
we have to consider the motion of a superparticle in superspace.
The superparticle in full superspace has a fermionic gauge symmetry
called $\kappa$-symmetry \cite{Siegel:1983hh}.

As noticed in ref.~\cite{Witten:1985nt}, for $\mathcal{N}=1$
super-Yang-Mills in ten dimensions, the translations in the fermionic
directions of the fat lines are $\kappa$-symmetry transformations.
Here we redo a similar analysis for $\mathcal{N}=4$ super-Yang-Mills
in four dimensions.  This could be done by dimensionally reducing
the $D=10$, $\mathcal{N}=1$ analysis, but we will redo it from scratch instead.

Let us now write down the worldline superparticle action with $\mathcal{N}=4$ supersymmetry.
According to \eqref{eq:interval}
the supercovariant momentum reads
\begin{equation}
  \pi = \dot x + i \theta \dot{\bar{\theta}}
- i \dot\theta\bar{\theta}.
\end{equation}
Then, the worldline superparticle action is ($g$ is the worldline einbein)
\begin{equation}
  \label{eq:superparticle}
  S = \sfrac{1}{2} \int d \tau \,g\, \pi^2.
\end{equation}  
This action is manifestly superconformal 
invariant since the momentum squared transforms homogeneously 
under inversions, by a factor which can be absorbed by the einbein $g$.

It is easy to show that the constraints in eq.~\eqref{eq:nullinterval} 
follow from the equations of motion of the action~\eqref{eq:superparticle} 
and that the solution in~\eqref{eq:fatcoords} 
is the general solution of these equations of motion.

The worldline reparametrisations are gauge symmetries which 
can be fixed by setting $g$ to be constant 
(but this gauge condition is not preserved by superconformal transformations).

Now we can explain in a different way why a straight light-like 
line in full superspace is not preserved by superconformal transformations.  
In the language of eq.~\eqref{eq:fatcoords}, if we take $\sigma$ and $\bar{\sigma}$ 
to be linear in $\tau$, after a superconformal transformation 
we need need to perform a compensating worldline reparametrisation $\tau \to \tau'(\tau)$ 
to preserve the gauge $g = \text{const}$.  
Since this reparametrisation is not linear in $\tau$, 
the odd coordinates $\sigma$ and $\bar{\sigma}$ 
will not be linear in the new worldline coordinate $\tau'$.

The action in eq.~\eqref{eq:superparticle} is also invariant 
under a local $\kappa$-symmetry which acts as
\[
  \delta \theta = \pi \bar{\kappa}, \quad
  \delta \bar{\theta} = \kappa \pi,\quad
  \delta x = -i \theta \kappa \pi  + i \pi\kappa \bar{\theta},
\quad
  \delta g = -2 i g \Tr \bigbrk{\bar{\kappa} \dot{\bar{\theta}}- \dot\theta\kappa}.
\]

\begin{figure}\centering
\includegraphics{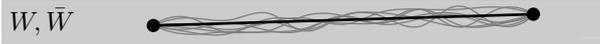}
\caption{All curves on a fat null line are physically equivalent;
they define the equivalent superparticle trajectories and equivalent Wilson lines.}
\label{fig:LineConnect}
\end{figure}

Now, we act with $\kappa$-symmetry on a superparticle
at the point $(x, \theta, \bar{\theta})$
whose supermomentum $\pi$ is light-like,
i.e.\ $\pi = \lambda \bar{\lambda}$.
We obtain
\[\label{eq:kappa}
\delta x = i \lambda \sigma \bar{\theta} - i \theta \bar{\sigma} \bar{\lambda},
\qquad
\delta\theta = \lambda \sigma,
\qquad
\delta\bar{\theta} = \bar{\sigma} \bar{\lambda},
\]
where we introduced the abbreviations $\sigma = \bar{\lambda} \bar{\kappa}$, $\bar{\sigma} = \kappa \lambda$.
Comparison to \eqref{eq:fatcoords} shows that $\kappa$-symmetry
can shift the point along any of the fermionic directions of a fat null line.
This implies that all paths along this fat null line
should be considered physically equivalent because
$\kappa$-symmetry is a gauge symmetry, cf.\ \figref{fig:LineConnect}.

So we see that the $\kappa$-symmetry transformations generate
a $(0\lvert 8)$-dimensional space
(the quantities $\sigma$ and $\bar{\sigma}$ are complex conjugate fermionic coordinates
with four complex dimensions, or eight real dimensions).
Here we notice a reduction by half of the number of transformation parameters;
we started with 16 real degrees of freedom in $\kappa$ and $\bar{\kappa}$,
but the latter only appear in the combinations $\sigma$ and $\bar\sigma$,
in which half of the degrees of freedom were projected out.

Using the $\kappa$-symmetry transformations in eq.~\eqref{eq:kappa}, 
we can compute its action on the twistor variables defined in eqs.~\eqref{eq:twistor,eq:twistorconj}.  
It is very easy to see that the twistor variables are invariant under $\kappa$ symmetry transformations.  
This was first noticed in ref.~\cite{Shirafuji:1983zd}.

\subsection{Yang--Mills Connection}
\label{sec:SYM}

Next we will discuss the implications
of the fatness of null lines for Wilson lines in
$\superN=4$ SYM \cite{Brink:1976bc,Gliozzi:1976qd}.
As a first step we will review the $\superN=4$ superspace
formulation \cite{Sohnius:1978wk},
in the next section we will apply it to Wilson lines.

To define Yang--Mills theory, we introduce a gauge connection one-form $A$
on superspace.
A generic gauge connection would have way too many degrees of freedom
as compared to the fields of $\superN=4$ supersymmetric Yang--Mills theory.
Therefore one must impose constraints on $A$
which is achieved by forcing
some components of the associated field strength $F=dA+A^2$ to zero.
This in turn not only reduces to the desired field content,
but also enforces the equations of motion.

Before we continue, let us briefly discuss differential forms on superspace.
It is convenient to express the components of differential forms in terms of
the superspace vielbein
$(d\theta^{\beta a},d\bar\theta_{b}{}^{\dot\alpha},e^{\beta\dot\alpha})$
where, according to \eqref{eq:interval},%
\footnote{The differential operator $d$ obeys the same statistics as fermions.
Consequently, $d\theta$ is bosonic.}
\[\label{eq:vielbein}
e=dx-id\theta\bar\theta-i\theta d\bar\theta.
\]
In particular, the exterior derivative $d$ can be expanded in this basis
\[
d
:=d\theta^{\beta a} \frac{\partial}{\partial\theta^{\beta a}}
+d\bar\theta_{b}{}^{\dot\alpha}\frac{\partial}{\partial \bar\theta_{b}{}^{\dot{\alpha}}}
+dx^{\beta\dot\alpha} \frac{\partial}{\partial x^{\beta\dot\alpha}}
=d\theta^{\beta a} D_{a\beta}
-d\bar\theta_{b}{}^{\dot\alpha}\bar D_{\dot{\alpha}}{}^b
+e^{\beta\dot\alpha} \partial_{\dot\alpha\beta}.
\]
Comparison of the definition of $d$
gives rise to supersymmetry covariant derivatives
\begin{equation}
  D_{a \beta} = \frac{\partial}{\partial \theta^{\beta a}}
+ i \bar{\theta}_a{}^{\dot{\gamma}} \frac \partial {\partial x^{\beta \dot{\gamma}}}
\,, \qquad
\bar{D}_{\dot{\alpha}}{}^b = -\frac \partial {\partial \bar{\theta}_b{}^{\dot{\alpha}}}
- i \theta^{a \gamma} \frac \partial {\partial x^{\gamma \dot{\alpha}}}
\,,\qquad
\partial_{\dot\alpha\beta}=
\frac{\partial}{\partial x^{\beta\dot\alpha}}
\,.
\end{equation}
which satisfy the $\superN=4$ super-Poincar\'e algebra with a flipped sign
\begin{equation}
\bigacomm{D_{a\beta}}{D_{c\delta}}
= \bigacomm{\bar{D}_{\dot{\alpha}}{}^b}{\bar{D}_{\dot{\gamma}}{}^d} = 0,
\qquad
\bigacomm{D_{a\beta}}{\bar{D}_{\dot{\gamma}}{}^d}
= -2 i \delta_a^d \partial_{\dot{\gamma}\beta}.
\end{equation}

The expansions of a generic gauge connection $A$
and its associated field strength $F=dA+A^2$ read
\begin{align}
\label{eq:AandF}
A=\mathord{}&d\theta^{\beta a} A_{a\beta}
-d\bar\theta_{b}{}^{\dot\alpha}\bar A_{\dot{\alpha}}{}^b
+e^{\beta\dot\alpha} A_{\dot\alpha\beta},
\nln
F=\mathord{}&
\half d\theta^{\beta a}  d\theta^{\delta c} F_{a\beta c\delta}
+\half d\bar\theta_b{}^{\dot\alpha}  d\bar\theta_{d}{}^{\dot\gamma} \bar F_{\dot\alpha}{}^b{}_{\dot\gamma}{}^d
-d\theta^{\beta a}  d\bar\theta_{d}{}^{\dot\gamma} F_{a\beta\dot\gamma}{}^d
\nln&
-d\theta^{\beta a}  e^{\delta \dot\gamma} F_{a\beta \dot\gamma\delta}
-e^{\beta\dot\alpha}  d\bar\theta_{d}{}^{\dot\gamma} \bar F_{\dot\alpha \beta\dot\gamma}{}^d
+\half e^{\beta\dot\alpha}  e^{\delta\dot\gamma} F_{\dot\alpha \beta\dot\gamma \delta}.
\end{align}
We use the connection to define a gauge covariant derivative
$\mathcal{D}=d+A$, or in components
\begin{equation}
  \mathcal{D}_{a\beta} = D_{a\beta} + A_{a\beta}, \qquad
  \bar{\mathcal{D}}_{\dot{\alpha}}{}^b
= \bar{D}_{\dot{\alpha}}{}^b + \bar{A}_{\dot{\alpha}}{}^b, \qquad
  \mathcal{D}_{\dot\alpha \beta}
= \partial_{\dot\alpha \beta} + A_{\dot\alpha \beta}.
\end{equation}
The components of the gauge-covariant field strength read
\begin{align}
F_{a\beta c\delta}
 &= \bigacomm{\mathcal{D}_{a \beta}}{\mathcal{D}_{c \delta}},
&
\bar F_{\dot\alpha}{}^b{}_{\dot{\gamma}}{}^{d}
&= \bigacomm{\bar{\mathcal{D}}_{\dot{\alpha}}{}^b}{\bar{\mathcal{D}}_{\dot{\gamma}}{}^d},
&
F_{a\beta\dot\gamma}{}^d
&= \bigacomm{\mathcal{D}_{a\beta}}{\bar{\mathcal{D}}_{\dot{\gamma}}{}^d}
+ 2 i \delta_a^d \mathcal{D}_{\beta \dot{\gamma}},
\nln
F_{a\beta \dot\gamma\delta}
 &= \bigcomm{\mathcal{D}_{a \beta}}{\mathcal{D}_{\dot\gamma \delta}},
&
\bar F_{\dot\alpha \beta \dot{\gamma}}{}^{d}
&= \bigcomm{\mathcal{D}_{\dot{\alpha}\beta}}{\bar{\mathcal{D}}_{\dot{\gamma}}{}^d},
&
F_{\dot\alpha\beta\dot\gamma\delta}
&= \bigcomm{\mathcal{D}_{\dot\alpha\beta}}{\mathcal{D}_{\dot{\gamma}\delta}}.
\end{align}

The constraint to reduce the connection $A$
to the field content of $\superN=4$ SYM
is imposed via the lowest components of the field strength $F$
\begin{equation}\label{eq:Fconstraint}
F_{a\beta c\delta}
 = \varepsilon_{\beta \delta} \bar\Phi_{a c},
\qquad
\bar F_{\dot\alpha}{}^b{}_{\dot{\gamma}}{}^{d}
= \varepsilon_{\dot{\alpha} \dot{\gamma}} \Phi^{bd},
\qquad
F_{a\beta\dot\gamma}{}^{d} = 0,
\qquad
\Phi^{ab}=\half e^{i\alpha}\varepsilon^{abcd}\bar\Phi_{cd}.
\end{equation}
Here the expansion of superfields $\Phi$ and $\bar \Phi$
in terms of fermionic coordinates
contains the scalars of $\superN=4$ SYM
as lowest components.
The phase $\alpha$ in the relation between $\Phi$ and $\bar\Phi$ has
no physical significance and we can safely set it to zero.
The Bianchi identities $dF+FA+AF=0$ then fix all the remaining
higher components of $F$, in particular
\[\label{eq:Fhigher}
F_{a\beta \dot\gamma\delta}
=\varepsilon_{\beta \delta} \bar\Psi_{a\dot\gamma},
\qquad
\bar F_{\dot\alpha \beta \dot{\gamma}}{}^{d}
= \varepsilon_{\dot{\alpha} \dot{\gamma}} \Psi_\beta{}^{d},
\qquad
F_{\dot\alpha\beta \dot\gamma\delta}
=
\varepsilon_{\beta \delta}
\bar\Gamma_{\dot\alpha\dot\gamma}
+
\varepsilon_{\dot\alpha\dot\gamma}
\Gamma_{\beta\delta},
\]
where the new superfields are given as derivatives of $\Phi$ and $\bar\Phi$
\begin{align}
\bar\Psi_{a\dot\gamma}
&=-\sfrac{i}{6} \bigcomm{\bar{\mathcal{D}}_{\dot\gamma}{}^e}{\bar\Phi_{ae}},
&
\Psi_\beta{}^{d}
&=-\sfrac{i}{6} \bigcomm{\mathcal{D}_{e\beta}}{\Phi^{de}},
\nln
\bar\Gamma_{\dot\alpha\dot\gamma}
&=\sfrac{1}{48}\bigacomm{\bar{\mathcal{D}}_{\dot\alpha}{}^e}{\comm{\bar{\mathcal{D}}_{\dot\gamma}{}^f}{\bar\Phi_{ef}}},
&
\Gamma_{\beta\delta}
&=\sfrac{1}{48}\bigacomm{\mathcal{D}_{e\beta}}{\comm{\mathcal{D}_{f\delta}}{\Phi^{ef}}}.
\end{align}
Furthermore, they imply a set of differential constraints
on the fields $\Phi$ and $\bar \Phi$
\begin{align}
0&=
3\bigcomm{\bar{\mathcal{D}}_{\dot\gamma}{}^d}{\bar\Phi_{ab}}
+\delta^d_a \bigcomm{\bar{\mathcal{D}}_{\dot\gamma}{}^e}{\bar\Phi_{be}}
-\delta^d_b \bigcomm{\bar{\mathcal{D}}_{\dot\gamma}{}^e}{\bar\Phi_{ae}},
\nln
0&=
\bigcomm{\mathcal{D}_{\delta c}}{\bar\Phi_{ab}}
+\bigcomm{\mathcal{D}_{\delta b}}{\bar\Phi_{ac}},
\nln
0&=
\bigacomm{\mathcal{D}_{\gamma d}}{\comm{\bar{\mathcal{D}}_{\dot\epsilon}{}^d}{\bar\Phi_{ab}}}
-\bigacomm{\bar{\mathcal{D}}_{\dot\epsilon}{}^d}{\comm{\mathcal{D}_{\gamma d}}{\bar\Phi_{ab}}}.
\end{align}
These equations are equivalent to the equations of motion
of $\superN=4$ SYM.

\subsection{Wilson Loop on a Fat Null Polygon}
\label{sec:wilsonloop}

Now consider a fat null polygon of dimension $1|8$.
To define a Wilson loop we need to embed a curve of dimension $1|0$
into the fat polygon.
It must pass through the vertices, but precisely which path
should it take on the fat null lines?
As for the trajectory of the superparticle and $\kappa$-symmetry,
the choice of curve within a null line does not matter \cite{Witten:1978xx,Witten:1985nt,Ooguri:2000ps}.
The crucial insight is that the Yang--Mills superspace
connection $A$ is flat on fat null lines.
This in turn implies the gauge field constraints 
and therefore the equations of motion.

On-shell the field strength $F$
reads \eqref{eq:AandF,eq:Fconstraint,eq:Fhigher}
\begin{align}\label{eq:Fonshell}
F=\mathord{}&
-\half \Tr (d\theta^{\trans}\varepsilon d\theta\,\bar\Phi)
+      \Tr (e^{\trans}\varepsilon d\theta\,\bar\Psi)
+\half \Tr (e^\trans\varepsilon e\,\bar\Gamma)
\nln&
-\half \Tr (d\bar\theta\varepsilon d\bar\theta^\trans\,\Phi)
+      \Tr (d\bar\theta \varepsilon e^\trans\,\Psi)
+\half \Tr (e\varepsilon e^\trans\,\Gamma).
\end{align}

On the fat null line \eqref{eq:fatcoords} the vielbein \eqref{eq:vielbein} read
\[
e=
(d\tau
-id\sigma\bar\sigma
-i\sigma d\bar\sigma
)
\lambda\bar\lambda,
\qquad
d\theta=\lambda d\sigma,
\qquad
d\bar\theta=d\bar\sigma\bar\lambda.
\]
As they are all collinear to $\lambda$ and/or $\bar\lambda$,
all the combinations in \eqref{eq:Fonshell} vanish
irrespectively of all the constituent fields,
and $F=0$ on fat null lines.
Conversely, the requirement $F=0$ on all fat null lines
essentially forces $F$ to be of the form \eqref{eq:Fonshell},
and thus the connection has to obey the constraints of $\superN=4$ SYM
along with the implied equations of motion.

\begin{figure}\centering
\includegraphics{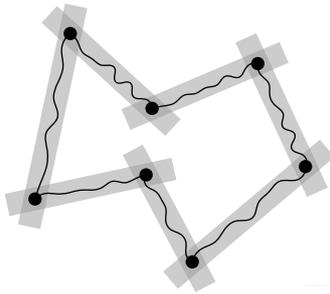}
\caption{A fat polygon with an embedded Wilson loop.}
\label{fig:FatPoly}
\end{figure}

There is no need to specify further the fermionic coordinates
of the Wilson line, as long as they reside fully within the fat null line.
Any section $\sigma(\tau),\bar\sigma(\tau)$ of the fat null line
yields the an equivalent Wilson line,
cf.\ \figref{fig:LineConnect}.
The latter depends only on the starting and end point,
which are two consecutive vertices by definition.
Altogether the fat polygon defines a family of
equivalent contours for a Wilson loop, cf.\ \figref{fig:FatPoly}.

Of course, in the quantum theory the Wilson loop needs to be regularised
for a proper definition. For bosonic Wilson loops dimensional reduction
is sufficient to regularise the UV divergences.
Conversely, for Wilson loops in superspace, 
the integrability condition on fat null lines is crucial, 
but it depends on the equations of motions
which are susceptible to UV quantum effects
\cite{Ooguri:2000ps,arXiv:1103.3008}.
Hence the Wilson loop expectation values have to
be regularised and quantised carefully. 
At least for the leading perturbative correction
at one loop it is possible to extract the result 
with only few complications as will be shown in the 
companion paper \cite{Beisert:2012xx}.

Suppose we consider Wilson loops without cusps, or we try to smooth
out the cusps to regularize the answer.  Then we can use
kappa symmetry to locally gauge away the dependence on the odd variables of the fat lines.  
In contrast, if the Wilson loops have cusps, 
the odd variables cannot be gauged away at the vertices because 
there the odd directions of the fat lines intersect transversely.  
It follows that the dependence on the odd variables 
is of a very different nature in the case with cusps and without cusps.

\section{Conclusions}
\label{sec:concl}

In this paper we have
detailed the definition of null polygons in full superspace.

We have presented three descriptions, in terms of the vertices,
in terms of spinor helicity variables,
and in terms of ambitwistor variables (\secref{sec:polygons}).
These generalise the analogous parametrisations
which were previously proposed for null polygons in bosonic spacetime and chiral superspace.
Importantly, they transform nicely under the full superconformal group,
and all of them are perfectly well-defined
in real spacetime with proper Minkowski signature.

A curiosity of the polygon's edges is that they are necessarily fat;
in addition to one bosonic coordinate, they have 8 fermionic coordinates
(\secref{sec:fat}).
Reassuringly, the fatness does not matter much
because all curves are physically equivalent in $\superN=4$ SYM theory
(\secref{sec:curves}).
We have also commented on the geometrical picture of the null polygon
in (real) ambitwistor space where it forms a dual polygon.

\medskip

Returning to the duality between planar scattering amplitudes
and null polygonal Wilson loops,
one may wonder how far it applies to our Wilson loop.
The picture we have obtained, however,
gives hints that the duality does not extend to full superspace.

Firstly, the segments are now parametrised by $(\lambda,\bar\lambda,\eta,\bar\eta)$
rather than $(\lambda,\bar\lambda,\eta)$. The additional four $\bar\eta$'s
suggests that a dual particle would have 16 times as many on-shell degrees of freedom.
From a physical point of view this does not make sense.

The identification with the momenta of particles bears another problem:
On the one hand, we might identify the bosonic momentum $p_k$
with the superspace interval $x_{k,k+1}$.
This is a null vector as it should for an on-shell particle.
Unfortunately, the intervals $x_{k,k+1}$ in
\eqref{eq:interval} do not sum up to zero due to the fermionic contributions.
Therefore the corresponding amplitude would violate momentum conservation.
On the other hand, we might identify $p_k$ with $x_{k+1}-x_k$.
Then the sum of momenta vanishes nicely. Instead, $p_k$ does not square
to zero anymore due to the fermionic contributions.
Hence, the corresponding particles cannot be massless.%
\footnote{This matches nicely with the minimal length $256$ for a massive supermultiplet.}

Even if these conflicts prevent a direct duality, it does not mean that
the Wilson loop in full superspace is useless for the duality.
For instance, it is the only kind of Wilson loop to which the
full set of superconformal transformations apply (up to anomalies at loop level).
The extended set of symmetries may make it easier to construct,
in particular in view of integrability in the form of Yangian symmetry \cite{Drummond:2009fd}.
Once constructed, we can set $\bar\eta=0$, and recover the Wilson loop
in chiral superspace which appears in the
duality to the complete scattering amplitude \cite{arXiv:1009.2225,arXiv:1010.1167}.%
\footnote{Note that scattering amplitudes
are not intrinsically chiral as their Wilson loop counterparts.
In contradistinction to chiral Wilson loops,
the full set of superconformal transformations applies to the S-matrix.
This apparent discrepancy does not spoil the duality because the
MHV-tree factor of the duality can compensate the mismatch.
The latter is also the reason for the absence of collinear anomalies
in Wilson loops which had to be cured by a deformed superconformal representation in \cite{Bargheer:2009qu}.}
Moreover, the supersymmetric anomaly of the chiral Wilson loop is also encoded
into the full Wilson loop \cite{CaronHuot:2011ky,CaronHuot:2011kk,Bullimore:2011kg}.

We would like to point out that the full superspace approach can indeed be useful
for the complete duality between null polygonal Wilson loops 
and null correlation functions of local operators 
\cite{arXiv:1007.3246, arXiv:1007.3243, arXiv:1009.2488, arXiv:1103.3714, arXiv:1103.4353, arXiv:1108.3557}
because both sets of observables are naturally defined on this superspace.

For superconformal theories in odd dimensions (like the ABJM theory~\cite{Aharony:2008ug}) 
it is not possible to construct a chiral version of the supersymmetric Wilson loop.  
However, most of the discussion we presented still applies.  
It is not completely clear what version of superspace would be the best suited in this case; 
so far most of the descriptions have been done in $\mathcal{N}=2$ superspace~\cite{Benna:2008zy}, 
but an $\mathcal{N}=2$ supermultiplet does not contain all the fields in the theory.   
It would be natural to use a gauge connection which contains all the physical fields of the theory, 
like for $\mathcal{N}=4$ super Yang--Mills.

\subsection{Acknowledgements}

We are grateful to
Simon Caron-Huot,
Song He,
Tristan McLoughlin,
David Mesterhazy,
Matteo Rosso
and
Burkhard Schwab
for discussions.
We are indebted to David Skinner for initial collaboration,
for sharing many of his insights with us and for comments on the manuscript.
During the preparation of this work we have benefited from the hospitality
of Kavli Institute for Theoretical Physics during the
``Harmony of the Scattering Amplitudes'' program and of Perimeter Institute
during the ``Integrability in Gauge and String Theory'' conference.

The work of NB is partially supported 
by grant no.\ 200021-137616 from the Swiss National Science Foundation
and by grant no.\ 962 by the German-Israeli Foundation (GIF).  
While he was at Brown University, CV was supported by the US Department of
Energy under contract DE-FG02-91ER40688 and the US National Science
Foundation under grant PHY-0643150.
The research at Kavli Institute for Theoretical Physics was supported
in part by the U.S.\ National Science Foundation under grant no.\ NSF PHY05-51164.

\phantomsection
\addcontentsline{toc}{section}{\refname}
\bibliography{superwilsongeo}
\bibliographystyle{nb}

\end{document}